\documentclass[prc,nofootinbib,showpacs,showkeys,reprint]{revtex4-1}
\usepackage{graphicx}
\usepackage{amssymb,amsmath,amstext,amsthm,amsfonts}

\newcommand{\dd}{\mathrm{d}}

\begin{document}

\today
\title{Mass dependence and isospin dependence of short-range correlated pairs}

\author{U. Mosel}
\email[Contact e-mail: ]{mosel@physik.uni-giessen.de}
\affiliation{Institut f\"ur Theoretische Physik, Universit\"at Giessen, Giessen, Germany}
\author{K. Gallmeister}
\affiliation{Institut f\"ur Theoretische Physik, Johann Wolfgang Goethe-Universit\"at Frankfurt, Frankfurt a.\ M., Germany}

\begin{abstract}
The target-mass number dependence of nucleon-nucleon pairs with short-range correlations is explored in a physically transparent geometrical model within a zero-range approximation. The observed $A$ dependence of 2-nucleon ejection cross sections in $(e,e')$ reactions is found to reflect the mass dependence of nuclear density distributions. A parametrization of this $A$ dependence is given. The $A$ dependence of proton-proton vs.\ proton-neutron pairs relative to $^{12}$C is also analyzed in this model. It can be understood using simple combinatorics without any additional isospin dependence.
\end{abstract}

\maketitle

\section{Introduction}
Recently the mass ($A$) dependence and quantum numbers of short-range correlated (src) pairs were extracted from $A(e,e'pp)$ and $A(e,e'pn)$ reactions \cite{Colle:2015ena,Subedi:2008zz}. There the observed number of proton-proton and proton-neutron pairs was used to constrain the number of initial-state pairs, their quantum numbers and their target mass-dependence. In a series of publications the Ghent group had explored these properties theoretically \cite{Vanhalst:2011es,Vanhalst:2012ur,Vanhalst:2011es,Vanhalst:2014cqa}. Their theoretical method was based on applying a $NN$ correlation operator, that contains essential features of the $NN$ interaction, to a many-body wave function obtained from a harmonic oscillator potential. All the src effects are then contained in that correlation operator. The authors of Ref.\ \cite{Colle:2015ena} showed that the mass-dependence of the data on $pp$ and $pn$ production could indeed be understood in this theoretical framework by assuming a zero-range approximation (ZRA) for the two interacting nucleons. By comparing theoretical results for a nucleus with a large neutron excess, such as $^{208}$Pb, with data one could hope to gain information also on the isospin content of the src pairs.

It is the purpose of this short paper to point out that the observed $A$-dependence is a consequence of nuclear density distributions. We will also apply simple combinatorics to explore the $A$-dependence of proton-proton to proton-neutron ratios.

\section{Geometrical Model}
In general any interaction between nucleons depends on the probability to find a nucleon at position $\mathbf{r}$ and simultaneously another one at $\mathbf{r'}$ and is  $\propto \rho(\mathbf{r})\,\rho(\mathbf{r'}) g(\mathbf{r} - \mathbf{r}')$. For a zero-range interaction $g(\mathbf{r} - \mathbf{r}') \to \delta(\mathbf{r} - \mathbf{r}')$ the  average probability density for finding a pair is then given by
\begin{equation}     \label{PNN}
P_{NN} = \int \dd^3r\, \rho^2(\mathbf{r}) = A \,\langle \rho \rangle   ~,
\end{equation}
where $\rho(\mathbf{r})$ is the single-particle density for a nucleus with mass number $A$ and $\langle \rho \rangle$ is the average nucleon density. In the following we take $P_{NN}$ as a measure for the overall strength of the src. Assuming some generic (e.g.\ Woods-Saxon) density distribution one sees that $P_{NN}\to \rho_0^2 V = \rho_0 A$ for $A \to \infty$; here $\rho_0$ is the nuclear matter saturation density and $V$ the nuclear volume. Thus, for large mass numbers $A$, $P_{NN}$ approaches a linear dependence on $A$. For smaller values one expects a correction due to the nuclear surface that is governed by the width parameter in the density distribution.

\begin{figure}
\centering
\includegraphics[width=0.7\linewidth,angle=-90]{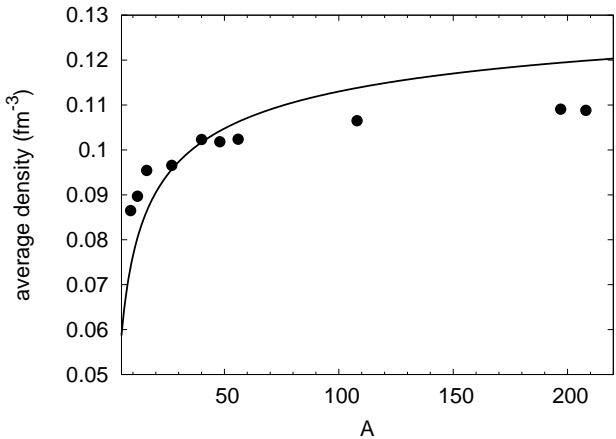}
\caption{Average density per nucleon $\langle \rho \rangle$ of Eq.\ (\ref{rhoav}) as a function of mass number $A$ (solid curve). The points are taken from Table 1 in \cite{Vanhalst:2014cqa}. For a shape comparison they were approximately scaled to the values of $\langle \rho \rangle$.}
\label{fig:P(A)}
\end{figure}

This is indeed borne out by computing the average density in (\ref{PNN}) using the experimental density distributions given in Ref.\ \cite{DeJager:1987qc}. Figure \ref{fig:P(A)}
shows a fit to this computed probability per nucleon. In a very good approximation the $A$-dependence of the average density for nuclei between $^{12}$C and $^{208}$Pb is described by
\begin{equation}  \label{rhoav}
\langle \rho \rangle (A) = 0.145 - 0.147 A^{-1/3} (\rm fm^{-3})  ~.
\end{equation}
The first constant term just gives the nuclear matter density $\rho_0$ and the second term represents a surface correction.

Also shown in Fig.\ \ref{fig:P(A)} are the values of the normalization of the correlated many-body wave function calculated in Ref.\ \cite{Vanhalst:2014cqa} which are a measure for the effect of the src operators acting on the harmonic oscillator ground state wave function. For a shape comparison these values were scaled down to the values of $\langle \rho \rangle(A)$. They follow qualitatively the behavior exhibited by the curve representing Eq.\ (\ref{rhoav}) with an $A$-dependence
$\label{NRyck}
\mathcal{N} = 0.127 - 0.0651 A^{-1/3} ~.
$
Because of the arbitrary normalization the main difference to the dependence given in Eq.\ (\ref{rhoav}) is in the surface term; we speculate that this difference could be connected with the unrealistic surface properties of harmonic oscillator wave functions used in Ref. \cite{Vanhalst:2014cqa}\footnote{We have found that the shape of the points can be very well described by using a Woods-Saxon density distribution with an unrealistically small surface parameter of a = 0.2 fm.}.

\section{Number of pairs}
The calculated $A$-dependence of $\langle \rho \rangle(A)$ (Fig.\ \ref{fig:P(A)}) is rather flat for heavier $A$; the different behavior for the lightest nuclei is due to the increasing relative importance of surface vs.\ volume effects with decreasing $A$. If we insert Eq.\ (\ref{rhoav}) into Eq.\ (\ref{PNN}), the geometrically predicted $A$-dependence of the pair-probability density is given by
\begin{equation}     \label{PNNA}
P_{NN}(A) = 0.145 A  - 0.147 A^{2/3} \quad (\rm fm^{-3})  ~.
\end{equation}

\begin{figure}[t]
\centering
\includegraphics[width=0.7\linewidth,angle=-90]{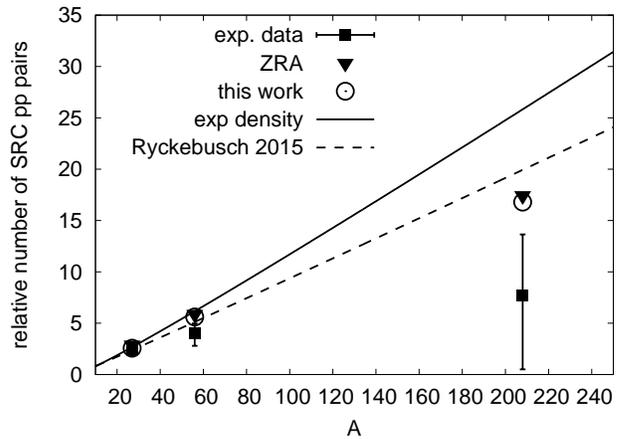}
\caption{Number of $pp$ src pairs relative to $^{12}$C. Solid line: $A$ times average density $A\,\langle \rho \rangle$ of Eq.\ (\ref{PNNA})  as a function of mass number $A$. Dashed line: mass-dependence fitted to results from Ref.\ \cite{Vanhalst:2014cqa}. Both curves are normalized to 1 for $^{12}$C. ZRA points (triangles) and data from Ref.\ \cite{Colle:2015ena}. Open circles: simple pair count (see text).  }
\label{fig:ppsrc}
\end{figure}

\begin{figure}[h]
\centering
\includegraphics[width=0.7\linewidth,angle=-90]{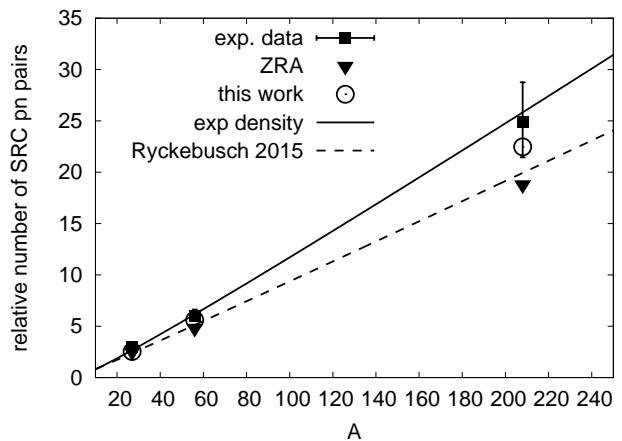}
\caption{Same as Fig.\ \ref{fig:ppsrc} for $pn$ pairs.}
\label{fig:pnsrc}
\end{figure}

In Figs.\ \ref{fig:ppsrc} and \ref{fig:pnsrc} we show the number of $NN$ pairs relative to $^{12}C$ both for the $A$-dependence derived from the average density (Eq.\ (\ref{PNNA})) and for that fitted to the explicitly counted number of zero range pairs obtained in Ref.\ \cite{Vanhalst:2014cqa}; the latter curve is very close to the curve labeled 'ZRA' in Fig.\ 3 of Ref.\ \cite{Colle:2015ena}.

Also shown in these figures are i) the data points extracted from $(e,e'pp)$ and $(e,e'pn)$ cross sections \cite{Colle:2015ena} and ii) the ZRA points given in \cite{Colle:2015ena}. Also shown are iii) the points obtained by multiplying the values of the solid curve (\ref{PNNA}) with the double ratios $Z(Z-1)/(A (A-1))/C_{pp}$ and $2\,Z\,N/(A(A-1))/C_{pn}$ for $pp$ and $pn$, respectively, i.e.\ by the simple combinatorial ratio for the presence of $pp$ or $pn$ pairs. Here $C_{pp} = 6 \times 5/(12\times 11) = 5/22$ and $C_{pn} = 2 \times 6 \times 6/(12 \times 11) = 6/11$ are the corresponding ratios for $^{12}$C; about 55\% of all pairs are $pn$ whereas 23 \% are $pp$ by combinatorics alone. Essentially the same combinatorial factor also appears if one starts from Eq.\ (\ref{PNN}) with $\rho_p = Z/A \,\rho$ and $\rho_n = N/A \,\rho$ for the proton and neutron densities.

Obviously, the overall $A$-dependence is reproduced quite well for the $pn$ pairs; for $pp$ pairs the measured $A$-dependence seems to be weaker than the calculated one. For the two light nuclei there is good agreement with the experimental data for both nucleon flavors, both for the ZRA points and the combinatorial points.

However, for the heavy nucleus $^{208}$Pb with its large neutron excess the experimental value for $pp$ is lower than both of these points (Fig.\ \ref{fig:ppsrc}).
A different behavior shows up for the $pn$ pairs in Fig.\ \ref{fig:pnsrc}. Now for $^{208}$Pb both the ZRA number and the combinatorial number are closer to the experimental point, with the combinatorial point at the lower end of the error bar.

\section{Conclusions}
In the preceding section we have shown that the mass dependence of data on the relative number of $pp$ and $pn$ pairs in ($e,e')$ reactions can be understood with a set of minimal assumptions. The fact that the $A$ dependence is close to linear just reflects the mass-dependence of the average density. This shows that the underlying mechanism is connected with very short range or zero range interactions.  The soft mass-dependence was one of the main conclusions of Ref.\ \cite{Colle:2015ena} and we verify it here. While a rather sophisticated theory and impressive apparatus of many-body theory were used in Refs.\ \cite{Vanhalst:2014cqa,Colle:2015ena} one could gain the impression of an inherent complexity of the physics of short range correlations \cite{Bertsch:1995ig}. We have shown here that the observed $A$-dependence is a simple consequence of nuclear geometry; this picture also explains the observed deviation from a strictly linear dependence on $A$ in terms of surface effects. Equation (\ref{PNNA}) gives a simple parametrization of this $A$-dependence.

We have also shown that the experimentally observed $A$-dependence of the numbers of $pp$ vs.\ $pn$ pairs is mostly determined by geometry and combinatorics combined. Any effects of a predominance of $pn$ processes over those of $pp$, which shows up in the reference nucleus $^{12}$C \cite{Subedi:2008zz}, do not appear in the target mass-dependence. For both nucleon flavors the $A$-dependence of the experimental values is nearly compatible with a simple statistical counting rule. If one really wanted to take the observed discrepancies for $^{208}$Pb seriously one would have to conclude that the experimental $pp$ process seems to be somewhat suppressed compared to simple counting whereas $pn$ is roughly described without any significant enhancement over the combinatorial result.

The arguments given here are independent of any particular interaction (electromagnetic or weak) and should thus be applicable also for neutrino-nucleus interactions. With the advent of LAr detectors a larger mass ($^{40}$Ar) is being explored than in most other experiments ($^{12}$C).
Based on Eq.\ (\ref{PNNA}) we thus predict a ratio of about 4.2 for the presence of short-range pairs in Ar vs.\ C. Conversely, the experimental determination of this ratio for the so-called 2p2h processes could give information on their effective range.

\acknowledgements
One of the authors (UM) acknowledges many helpful discussions with Jan Ryckebusch.

This work was partially supported by Deutsche Forschungsgemeinschaft (DFG) and the Helmholtz International Center for FAIR.

\bibliographystyle{apsrev4-1}
\bibliography{nuclear}
\end{document}